\documentclass[conference]{IEEEtran}
\IEEEoverridecommandlockouts
\usepackage{amsmath,amssymb,amsfonts}
\usepackage{algorithmic}
\usepackage{graphicx}
\usepackage{textcomp}
\usepackage{xcolor}
\usepackage{mathtools}
\usepackage{glossaries}
\usepackage{booktabs}
\usepackage[numbers]{natbib}
\usepackage[utf8]{inputenc}
\usepackage[T1]{fontenc}
\usepackage{tabularx}
\usepackage{url}

\begin{document}
\title{Efficient Deep Learning-based Estimation of the Vital Signs on Smartphones}

\author{

\IEEEauthorblockN{\textsuperscript{*} Taha Samavati\textsuperscript{†}, 
\textsuperscript{*} Mahdi Farvardin \textsuperscript{†},
Aboozar Ghaffari\textsuperscript{+}}

\IEEEauthorblockA{\textsuperscript{†}\textit{ Department of Computer Engineering, Iran University of Science and Technology, Tehran, Iran}\\
\textsuperscript{+}\textit{ Department of Electrical Engineering, Iran University of Science and Technology, Tehran, Iran}\\
taha\_samavati@alumni.iust.ac.ir\\
m\_farvardin@alumni.iust.ac.ir\\
aboozar\_ghaffari@iust.ac.ir}
\thanks{* Equal Contribution}
}

\maketitle



\begin{abstract}

With the increasing use of smartphones in our daily lives, these devices have become capable of performing many complex tasks. Concerning the need for continuous monitoring of vital signs, especially for the elderly or those with certain types of diseases, the development of algorithms that can estimate vital signs using smartphones has attracted researchers worldwide. In particular, researchers have been exploring ways to estimate vital signs, such as heart rate, oxygen saturation levels, and respiratory rate, using algorithms that can be run on smartphones. However, many of these algorithms require multiple pre-processing steps that might introduce some implementation overheads or require the design of a couple of hand-crafted stages to obtain an optimal result. To address this issue, this research proposes a novel end-to-end solution to mobile-based vital sign estimation using deep learning that eliminates the need for pre-processing. By using a fully convolutional architecture, the proposed model has much fewer parameters and less computational complexity compared to the architectures that use fully-connected layers as the prediction heads. This also reduces the risk of overfitting. Additionally, a public dataset for vital sign estimation, which includes 62 videos collected from 35 men and 27 women, is provided. Overall, the proposed end-to-end approach promises significantly improved efficiency and performance for on-device health monitoring on readily available consumer electronics.

\end{abstract}


\section{Introduction}
Vital signs need to be monitored regularly, especially in the elderly or individuals with certain medical disorders. Nowadays, anyone has a smartphone and uses it for a variety of tasks on a daily basis. Due to the rapid development of both hardware and software, smartphones can perform more complex tasks. Having mentioned these, one can take advantage of smartphones for estimating and monitoring vital signs with near-clinical accuracy. Heart rate (HR), Oxygen saturation level (SpO2), and Respiratory Rate (RR) are three of the key vital signs in the human body. Heart Rate is an important indicator of people's physiological state and needs to be measured in many circumstances, especially for healthcare and medical purposes. The oxygen saturation level in our bloodstream, also known as SpO2, shows how much oxygen is carried by the blood. Measuring SpO2 levels is essential as a major deviation from normal levels (typically more than 95\% at sea level) \cite{american2003american-b1} indicates a dangerous health condition or a sign of a serious disease (including COPD, Asthma, interstitial lung diseases, sequelae of tuberculosis, lung cancer, chronic obstructive pulmonary disease, and COVID-19, that can cause significant drops in SpO2 levels)\cite{vold2015low-b2, goyal2021room-b3}. Another vital sign that indicates how well our respiratory system functions is the respiratory rate. It refers to the number of breaths taken per minute and reflects the body's physiological need for oxygen. Typical respiratory rates in healthy adults at rest range from 12 to 20 breaths per minute \cite{chourpiliadis_physiology_2022}. An increased respiratory rate (tachypnea) may indicate conditions such as pneumonia, sepsis, congestive heart failure, anxiety disorders, while a decreased rate (bradypnea) may be associated with drug overdose, hypothermia or neurological issues \cite{MOWER1996593}. Therefore, monitoring respiratory rate provides a significant insight into lung function and potential health complications.

To predict the vital signs using optical measurement methods, one should first obtain the Photoplethysmography signal. Photoplethysmography (PPG) is an uncomplicated and inexpensive optical measurement method that is often used for heart rate monitoring purposes. PPG is a non-invasive technology that uses a light source and a photodetector at the surface of the skin to measure the volumetric variations of blood circulation \cite{castaneda2018review-b4}. The PPG signal can also be obtained by smartphones with both a camera and a flashlight. By covering the camera with a finger while its flashlight is on, one can obtain photo sequence captures and further process them to obtain the PPG signal. Various methods have been proposed for vital sign estimation. Some of these methods only rely on signal processing algorithms to extract the vital signs from the input PPG signal \cite{siddiqui2016pulse-b5}, \cite{ding2018measuring-b6}, \cite{nemcova2020monitoring-b7}. while some use both signal processing and deep learning methods \cite{ayesha2021heart-b8,reiss2019deep-b9}. The rest focus on developing end-to-end deep learning methods \cite{shyam2019ppgnet-b10}, \cite{kang2022transppg-b11}, \cite{yang2022assessment-b12}. These mentioned methods either need multiple pre-processing stages or have a high computational burden to be run on mid-range or low-end smartphones. 

In this article, a set of architectures for real-time heart rate, SpO2, and respiratory rate estimation on mobile devices are proposed. These methods estimate heart rate, SpO2, and respiratory rate in an end-to-end manner with only a normalization step integrated directly into the model, which makes the deployment process easier for mobile devices. Compared to prior proposed architectures that used fully connected layers following the feature extraction stage, fully convolutional architectures do not use dense connections after extracting features; hence they have fewer parameters while achieving better accuracy. As another contribution of this research, a public dataset of smartphone videos named MTHS is provided, containing extracted PPG signals from 62 distinct patients with their corresponding ground truth HRs and SpO2s. More detailed information about the dataset can be found in the Section \ref{datasets}.

The paper begins by examining related works on vital sign estimation, followed by a detailed description of the datasets used and the data collection process. The proposed efficient deep learning architectures are then introduced, along with the experimental results on benchmark datasets compared to state-of-the-art methods. The deployment of these methods on a smartphone application is also demonstrated. Ultimately, the paper concludes with a thoughtful discussion of key findings and future directions for research.

\section{Related Works}
\textbf{Heat Rate (HR) Estimation:} The SpaMA algorithm \cite{salehizadeh2015novel} is a multi-stage approach to reconstruct PPG signals and extract heart rate information from data corrupted by motion artifacts. It utilizes joint time-frequency analysis of the PPG and accelerometer data, identifies and removes motion artifact frequencies based on accelerometer spectra, tracks the true heart rate peak over time. It then reconstructs the PPG signal using the extracted heart rate amplitude, frequency, and phase information. Heart rate variability analysis is then performed on the reconstructed PPG and heart rate time series with motion artifacts removed.

In Research \cite{schack2017computationally}, the authors propose a computationally efficient method for extracting heart rate from PPG signals corrupted by motion artifacts using only basic signal processing techniques like correlation, spectral analysis, and frequency tracking. Their approach achieves state-of-the-art accuracy on benchmark datasets while significantly reducing computational demand compared to other methods.

Research \cite{reiss2019deep-b9} tends to estimate heart rate by taking FFT of single-channel time-series PPG along with 3-axis accelerometer motion signals and feeds the resulting four-channel signals to the proposed neural network. The input signal is clipped between 0-4 Hz to remove unwanted frequencies. The proposed model consists of  8 channel 1D convolution-max pool layers followed by a fully connected network. 

Research \cite{shyam2019ppgnet-b10} proposes an end-to-end deep learning model to estimate heart rate from PPG signals acquired by a wrist-worn device. The proposed method does not require any pre-processing steps or any motion data. Nevertheless, the proposed method achieves competitive results compared to those using motion signals. In this work, eight consecutive one-second PPG data with a sample rate of 125 Hz are fed to a set of convolutional layers and an LSTM layer in parallel. The produced feature vectors are then concatenated together and fed to another LSTM layer. Finally, a linear layer predicts the heart rate. Aside from the advantage of eliminating the need for pre-processing steps, due to the presence of LSTM layers, the model still has a relatively high computational complexity.

Research \cite{ayesha2021heart-b8} utilizes a convolutional-based neural network to estimate the heart rate from PPG signals acquired from smartphone captured videos. The input videos are first converted to a set of 3 channel 1D signals. These signals are the mean values of the frames' red, green, and blue channels. Multiple pre-processing algorithms were applied to the signals before feeding them to the network. These pre-processing steps include multiple steps such as denoising, applying moving average, and PCA. This method utilizes fully connected layers following the convolutional layers.

BeliefPPG \cite{bieri2023beliefppg} presents a novel machine learning approach for accurate heart rate estimation from photoplethysmography (PPG) signals. The authors model the evolution of heart rate as a hidden Markov process and use a neural network to derive a statistical distribution over possible heart rate values for each PPG signal window. They then apply belief propagation to refine the estimates temporally based on the distribution of heart rate changes. This provides a quantized probability distribution over heart rates that captures inherent uncertainty. Robust testing on 8 public PPG datasets with 3 cross-validation strategies demonstrates state-of-the-art heart rate estimation performance. The model is able to leverage temporal heart rate patterns to improve accuracy and provide well-calibrated confidence measures. 

\textbf{SpO2 Esitmation:}
Research \cite{ding2018measuring-b6} estimates spo2 from smartphone videos captured from finger-tips. After the pre-processing step, which includes motion removal, a convolutional neural network is proposed for SpO2 estimation, which has only two 1D convolutional layers with a large filter length followed by max pool and dropout. A pulse oximeter records ground truth heart rate and SpO2, while an iPhone 7 plus is used to record fingertip videos. The researchers conclude that the performance of their proposed method does not improve when increasing the frame rate above 30fps. They also suggest using Raw PPG instead of pre-processed (band-pass filtered) as it achieves the best performance.

Research \cite{hoffman2022smartphone-b14}, proposes a convolutional neural network followed by a fully connected layer. The model takes in the mean channel values of gain applied RGB channels and predicts the SpO2 values. The captured video, from which the mean signals are extracted, has a time length of three seconds with a sampling rate of 30 fps. However, the collected dataset has limited coverage of patients. The proposed model uses fully connected layers after the CNN layers, resulting in higher parameter count and over-fitting chance compared to the fully convolutional architectures.

Research \cite{koteska2022machine-b16} extracts seven features from the PPG signal using two python kits, namely HeartPy and Neurokit. The researchers use machine learning methods to estimate SpO2 levels. The best performing method was the Random Forest regressor, which has achieved an MAE of 1.45 on BIDMC's test set. Although it is not stated as an advantage in the original work, the random forest regressor can also be deployed on mobile devices after performing proper conversion. In this method multiple pre-processing methods were applied to the input signal which can be listed as a draw back for this work.

\textbf{Respiratory Rate (RR) Estimation:}
Research \cite{jarchi2018accelerometry} presents two methods to continuously estimate respiratory rate (RR) - using accelerometer signals or photoplethysmograph (PPG) waveforms collected from wearable sensors in a clinical setting. The accelerometer approach involves adaptive line enhancement and singular spectrum analysis to preprocess the signals and identify accurate RR frequency components. The PPG-based approach first selects PPG segments having a high Signal Quality Index (SQI > 0.85). Respiratory-induced intensity and amplitude modulation signals are then extracted from the PPG. Auto-regressive spectral analysis is applied to each modulation signal to identify RR-related peak frequencies. The RR estimates from the two modulations are fused through a threshold-based comparison. If the difference is less than 3 bpm, their average is output as the final RR, otherwise no estimate is made. Respiratory rate estimates from both accelerometer and PPG signals show strong agreement across different patients over extended durations of more than two days. As accelerometers are more convenient for ambulatory monitoring, demonstrating their ability to reliably estimate RR opens up new possibilities for continuous respiration monitoring after intensive care without needing PPG sensors.

In \cite{bian2020respiratory}, an approach based on residual network (ResNet \cite{he2016deep}) architecture using end-to-end deep learning is proposed to estimate RR from PPG. This method takes time-series PPG data as input and learns the rules through a training process that involves a synthetic PPG dataset that was created to overcome the insufficient data problem in deep learning. The inclusion of the synthetic dataset for training improved the performance of the deep learning model by 34\%. The final mean absolute error performance for the deep learning approach in estimating RR was 2.5±0.6 bpm using 5-fold cross-validation in two widely used public PPG datasets with reliable RR references.

In research \cite{aqajari2021end}, the researchers present an end-to-end pipeline for RR estimation using Cycle Generative Adversarial Networks (CycleGAN) to reconstruct respiratory signals from raw PPG signals. The results demonstrate a higher RR estimation accuracy of up to 2 times (mean absolute error of 1.9±0.3 using five-fold cross-validation) compared to the previous methods on the BIDMC dataset.

Research \cite{shuzan2023machine}, presents a machine learning-based method of estimating RR and SpO2. It first extracts a set of meaningful features from the input PPG signal. Nine feature selection algorithms were used in this study, and the best one was selected. The feature selection process was used to reduce the computational complexity and the possibility of overfitting. Nineteen models were trained for both RR and SpO2 separately, from which the most appropriate regression model was the Gaussian process regression model that outperformed all the other models for both RR and SpO2 estimation.

\section{Proposed Methods}
We propose a highly efficient real-time algorithm to estimate vital human body signs such as heart rate, SpO2, and RR on smartphones and mobile devices. The proposed deep learning method has fewer parameters than the previously proposed architectures for vital sign estimation.

\begin{figure*}[htbp]
    \centering
    \includegraphics[width=\textwidth]{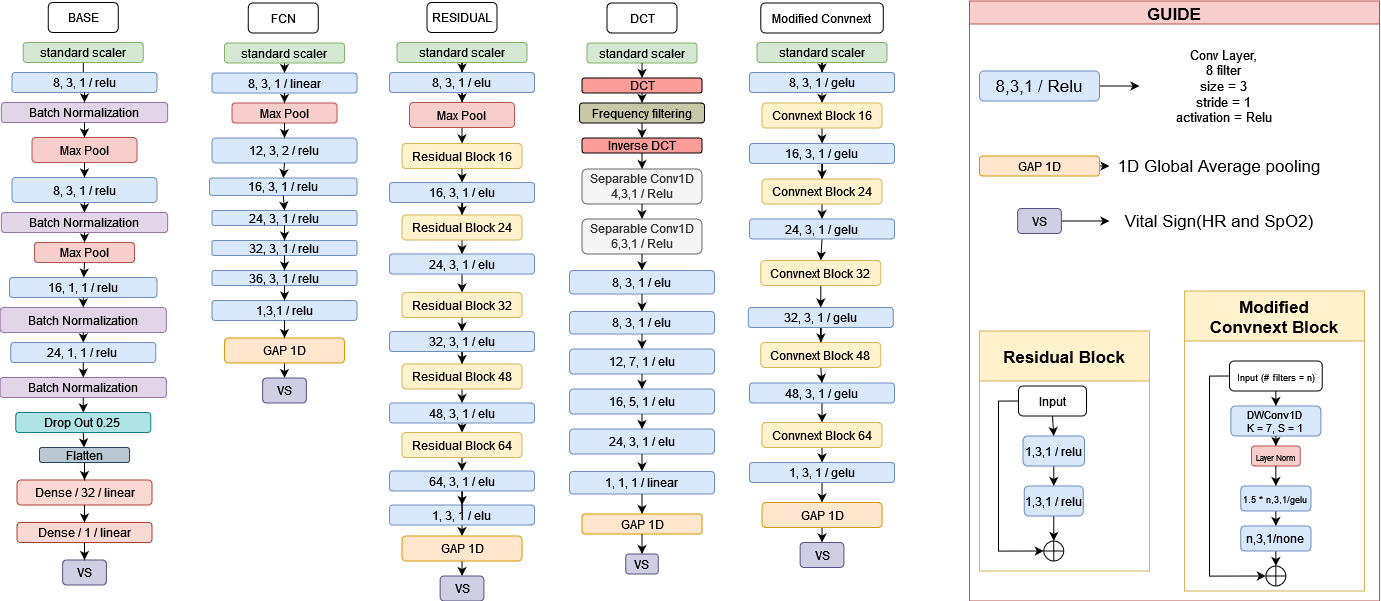}
    \caption{The proposed architectures for vital sign estimation, there are five different architectures provided. The Base model represents the commonly used architecture for this task, which is used as a baseline. The guide section provides detailed information about the blocks in the proposed architectures.}
    \label{fig:model_archs}
\end{figure*}

Given an image sequence or a video taken from fingertips, MEDVSE (Mobile-Efficient, Deep Learning-Based Vital Sign Estimation) can estimate HR, SpO2, and RR in real time. Our proposed network estimates vital signs in a fully convolutional manner. This results in a network with much fewer parameters than the conventional methods, which use a fully connected network after the convolutional layers. Moreover, having fewer parameters decreases the chance of over-fitting, which can be beneficial when having a small-sized dataset. The commonly used batch normalization layers are not used to further decrease the computational complexity. The proposed architectures eliminate the need for implementing pre-processing steps on the device's native development framework, which might introduce some overheads. We propose four types of architectures shown in Fig \ref{fig:model_archs}. The first architecture is a stack of 1D convolutional layers followed by a global average pooling (GAP) layer that produces the vital sign of interest (Heart rate, SpO2 or respiratory rate). The second one, Residual FCN, is deeper and adds residual connections to the normal convolutional layers. The third model, named DCT, first applies a Discrete Cosine Transformation (DCT) to the input PPG signal and filters un-related frequencies. Afterward, a series of convolutional layers learn the mapping to the desired vital sign. The transformation and coefficient filtering procedures are implemented in the model itself and can be directly deployed to other platforms without the need for re-implementation of these steps in other platform-specific programming languages. The last model, is a convnext-based architecture \cite{liu2022convnet}, in which a series of modified convnext blocks are stacked on top of each other. We simply remove the "Layer Scale" and "Drop Path" operations. In order to compare the performance of these methods against the conventional architectures, which use fully connected layers as the prediction heads, a base model is also implemented. The base model has a series of 1D convolutional layers followed by batch normalization as well as a fully connected network after the feature extraction phase. 

The model details such as parameter counts are listed in Table \ref{table:model-details}. The size of trained models is less than a MB on disk. With the exception of the Residual FCN model, all other models demand fewer than one Mega FLOPs. Despite the Residual FCN model having the highest FLOP count, its inference time still remains remarkably low while delivering 40 to 50 percent lower errors compared to the base model. For instance, on a mid-range device equipped with a single-core CPU clocked at 2.27 GHz, it completes a single forward run in a mere 0.03 milliseconds, showcasing the efficiency of the model even under demanding computational loads. The experiments clearly show that the best-performing architecture is the fully convolutional one. More detailed results are discussed in the Experimental Results section. 

\begin{table}[h]
\centering
\begin{center}
\caption{Details of the proposed networks.}
\label{table:model-details}
\begin{tabular}{llll}
\toprule
Model Name & Parameters  & FLOPs  & Size \\ \hline
Base                    & 29665 & 0.141 M  & 450 Kb     \\
FCN                   & 8037 & 0.485 M  & 170 Kb     \\
Residual FCN             & 53713 & 7.17 M & 800 Kb     \\
DCT                  & 3266 & 0.902 M & 150 Kb \\
Modified-Convnext            & 23017 & - & 345 Kb \\ 
\bottomrule
\end{tabular}
\end{center}
\end{table}

At inference time, the network estimates the vital signs from the fingertips. Therefore it first needs an image sequence or video; we obtain these by asking the patient to cover the smartphone's camera with his/her fingertip while the flashlight is on. Depending on the task, a single color channel is used for heart rate (the green), and double channels (green and blue) are used as input for SpO2 estimation. The image sequence is captured and processed simultaneously so that for each image in the input sequence, mean values for RGB channels are computed and listed in three arrays separately. After 10 seconds, the resulting mean RGB signals are extracted to be fed into MEDVSE for estimation of vital signs. As depicted in Figure \ref{fig:model_archs}, the input PPG signal is first transformed to have zero mean and unit standard deviation. The model inherently performs the latter process.

\section{Datasets}\label{datasets}
This section briefly discusses the datasets that are used to benchmark our proposed methods. In the following, the details of these datasets are explained.

\textbf{BIDMC\cite{pimentel2016toward-b15}: }This dataset contains PPG, impedance respiratory signal, and electrocardiogram (ECG) for 53 patients. Each recording is eight minutes long and also includes ground truth data for heart rate (HR), respiratory rate (RR), and blood oxygen saturation level (SpO2) sampled at 1Hz. It should be noted that, despite that the PPG signal was not acquired by a smartphone camera, our model can be trained on any type of input PPG signal.

\textbf{MTHS:} This research proposes a dataset from 62 patients (35 men and 27 women) that contains both HR and SpO2 labels sampled at 1Hz. The PPG signal is acquired at 30 FPS using a smartphone's camera. The data collection procedure is explained in Data Collection Section. 

\textbf{PPG-DaLiA\cite{ppgdalia2019}}: This multimodal dataset comprises physiological and motion data obtained from both a wrist-worn and a chest-worn device. The data were recorded from 15 subjects engaged in a diverse set of activities conducted under conditions closely resembling real-life scenarios. The dataset incorporates ECG data, serving as the ground truth for heart rate. Additionally, the dataset includes PPG and 3D-accelerometer data, which can be employed for heart rate estimation, effectively mitigating the impact of motion artifacts. The PPG signal is acquired at 64Hz and the corresponding labels are sampled at 0.5Hz.

\section{Data Collection} \label{data_collection_sec}

\begin{figure}[h]
    \centering
    \includegraphics[width=\textwidth/2]{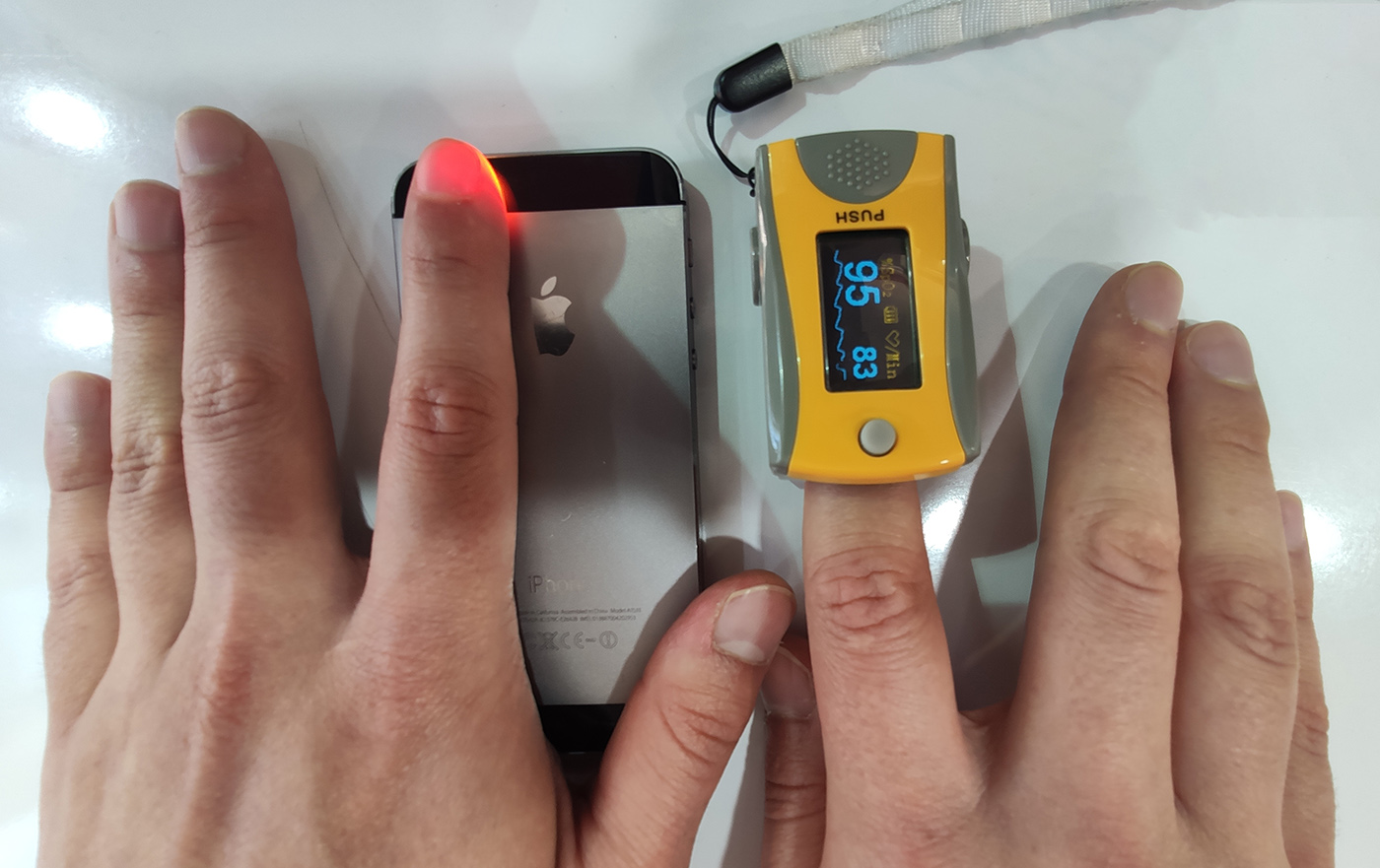}
    \caption{The setup for data collection.}
    \label{fig:setup-pics}
\end{figure}

\begin{figure}[h]
    \centering
    \includegraphics[width=0.5\textwidth]{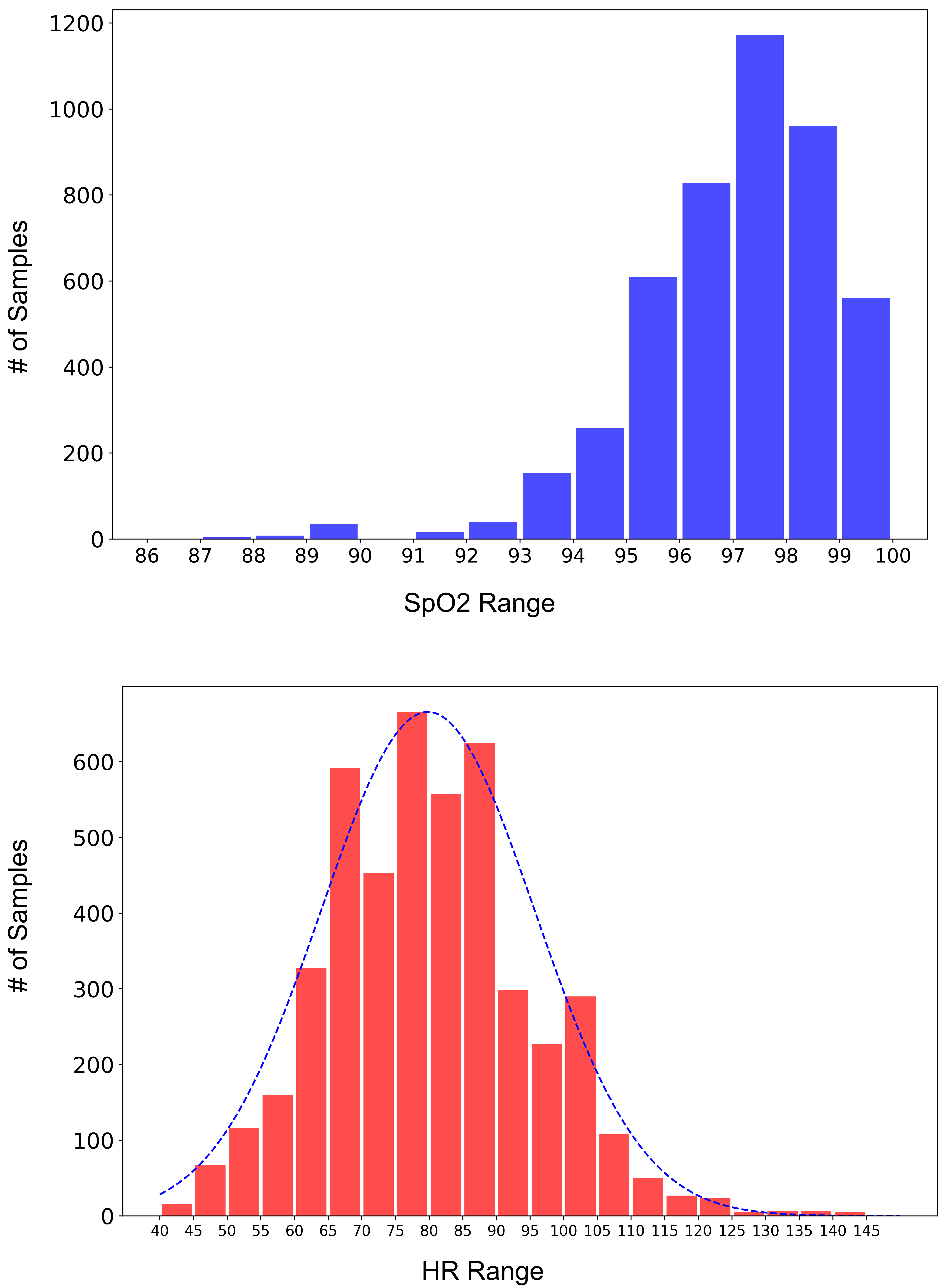}
    \caption{Heart rate and SpO2 distributions for the MTHS dataset.}
    \label{fig:BIDMC-hr-spo2-stats}
\end{figure}

In this dataset, the PPG is acquired using the phone's RGB camera. As there are very few publicly available datasets with a PPG signal acquired using a smartphone, we decided to collect such a dataset named MTHS. As described in the previous section, the MTHS dataset contains PPG signals with the sampling frequency of 30Hz obtained from 62 patients, including 35 men and 27 women. The ground truth data includes heart rate and oxygen saturation levels sampled at 1Hz. The HR and SpO2 measurement is obtained using a pulse oximeter (M70). An iPhone 5s was used to obtain the PPG recordings at 30 fps. The flashlight is kept on during the recording phase. The patients were asked to fully cover the camera and the flashlight with their fingertips. Figure \ref{fig:setup-pics} shows the data collection procedure. The dataset is openly available on Github\footnote{https://github.com/MahdiFarvardin/MEDVSE}. Figures \ref{fig:BIDMC-hr-spo2-stats} provides a statistical insight over the collected dataset.

\section{Experimental Results}
\label{section_exp_res}
In this section, The proposed methods are evaluated on a number of benchmark datasets, and the results are included. Furthermore, some visualizations of predictions and estimated values are provided.

The proposed architectures are named as BASE, FCN, Residual-FCN, DCT, and Modified ConvNext. The mentioned networks are trained and evaluated for 125 epochs on three datasets, including BIDMC, our proposed dataset called MTHS, and PPG-Dalia. The base model is based on common and older architectures in regression which use a CNN feature extractor followed by a feed-forward network. The FCN model has a fully convolutional architecture with about 4x lower parameters than the base model. This architecture makes over-fitting less likely to happen and highly efficient for deployment on mobile devices. As explained before, the DCT model has the same fully-convolutional architecture but works on the frequency domain. After taking the DCT of the input signal and filtering out DC coefficients and high-frequency coefficients that do not contain our desired frequencies for HR, a stack of 1D convolutions is applied to regress HR and SpO2 values. In the following, the performance of the proposed methods on each dataset is reported.

\subsection*{Experimental Results on BIDMC}
The proposed architectures are evaluated on BIDMC dataset by performing 10 fold cross validation over three different tasks of heart rate, SpO2 and RR estimation. According to the results in Table \ref{table:BIDMC-MAE-hr-spo2}, our proposed fully convolutional architecture with residual connections outperforms other architectures achieving a mean MAE of 2.20 on the BIDMC heart rate dataset. This proves the effectiveness of incorporating residual blocks in the model architecture. During the experiments on this dataset, it was found that the proposed FCN-DCT model performs better when removing the inverse DCT transformation. This is referred to as "w/o Inversion" in the results table. The SpO2 estimation results on BIDMC are also shown in Table \ref{table:BIDMC-MAE-hr-spo2}. These results demonstrate that the best performing architecture for SpO2 estimation is the DCT model. 

\begin{table}[b]
\centering
\caption{Heart rate and SpO2 estimation results on BIDMC. The mean and standard deviation of MAEs are reported with a 10-fold cross-validation setup.}
\label{table:BIDMC-MAE-hr-spo2}
\begin{tabular}{lcc}
\toprule
Model Name & Heart Rate (MAE) & SpO2 (MAE) \\ \hline
Base & 3.37 $\pm$ 1.37 & 3.76 $\pm$ 1.53 \\
FCN & 2.81 $\pm$ 1.55 & 3.57 $\pm$ 1.65 \\
Residual FCN & \textbf{2.20 $\pm$ 1.38} & 3.67 $\pm$ 1.76 \\
DCT (w/o Inversion) & 3.25 $\pm$ 0.72 & \textbf{3.14 $\pm$ 1.64} \\
Modified ConvNext & 2.38 $\pm$ 1.35 & 3.69 $\pm$ 1.68 \\ \bottomrule
\end{tabular}
\end{table}

\begin{figure}[b]
    \centering
    \includegraphics[width=0.5\textwidth]{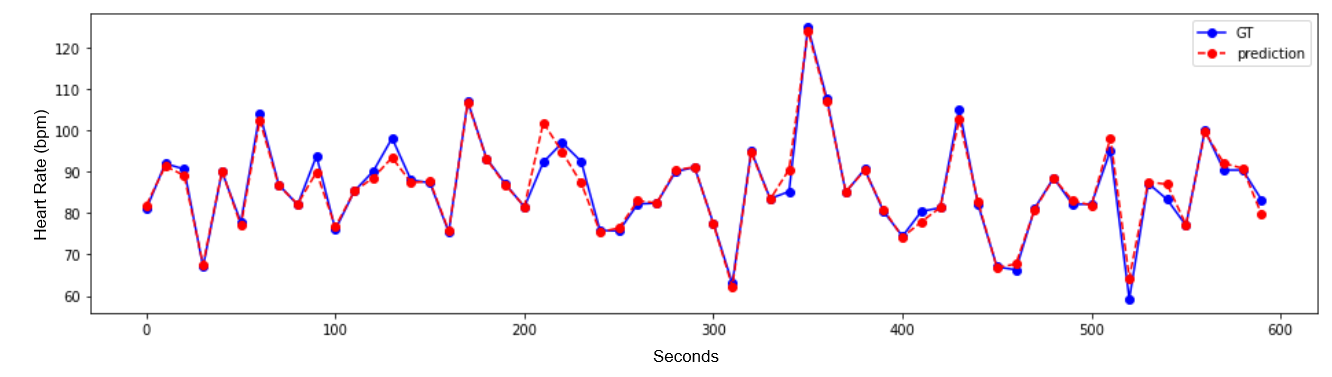}
    
    \caption{A plot of ground truth heart rate and its estimation by the best performing model (Residual FCN) on BIDMC dataset.}
    \label{fig:BIDMC-pred-hr}
\end{figure}

\begin{figure}[b]
    \centering
    \includegraphics[width=0.5\textwidth]{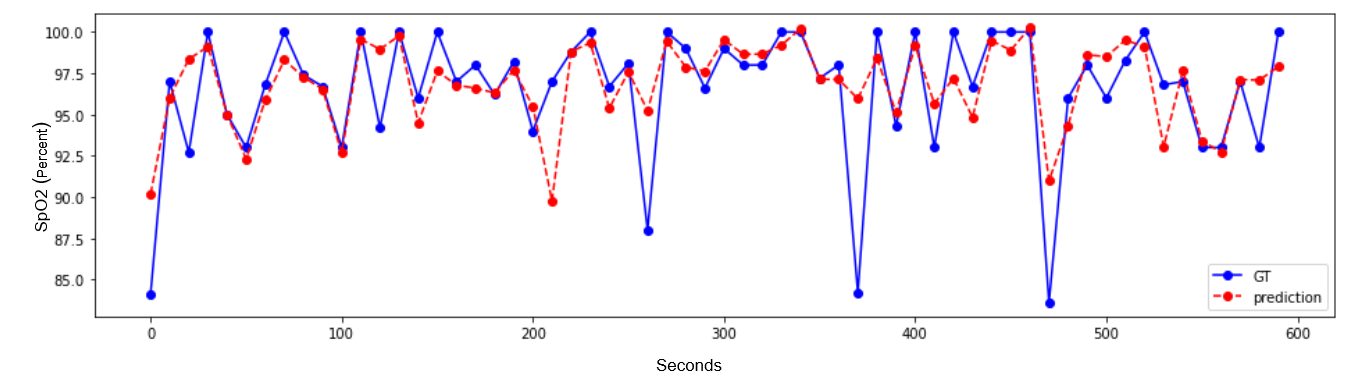}
    
    \caption{A plot of ground truth SpO2 and its estimation by the best performing model on the BIDMC dataset.}
    \label{fig:BIDMC-pred-spo2}
\end{figure}

Figures \ref{fig:BIDMC-pred-hr} and \ref{fig:BIDMC-pred-spo2}, provide two plots for both HR and SpO2 predictions on BIDMC test set with corresponding ground truth. Each plot contains 60 samples. To clarify more, each sample input is a ten-second signal and the corresponding label is the mean value of the target signal in those 10 seconds. the next sample has no overlapping and contains the next 10-second period's signal and the mean of their corresponding labels in that period. 

\subsection*{Experimental Results on MTHS}
Below is the mean and standard deviation of MAEs obtained by each of the proposed models. According to the results in Table \ref{table:MTHS-MAE-hr-spo2}, same as before, our proposed fully convolutional architecture with residual connections outperforms other architectures on the task of hear rate estimation, achieving a mean MAE of 7.64. The SpO2 estimation results are also provided in the Table \ref{table:MTHS-MAE-hr-spo2}. The best performing model (Residual FCN) achieves a mean MAE of 1.47 on the MTHS dataset. It is worth to note that the DCT model having only one tenth of the base model's parameters, achieves better results. However one might consider the extra FLOPs required by the DCT model.

\begin{table}[b]
\centering
\caption{Heart rate and SpO2 estimation results on MTHS dataset with 10-fold cross-validation. Mean and standard deviation of MAEs are reported.}
\label{table:MTHS-MAE-hr-spo2}
\begin{tabular}{lcc}
\toprule
Model Name & Heart Rate (MAE) & SpO2 (MAE) \\ \hline
Base & 11.20 $\pm$ 3.25 & 1.80 $\pm$ 0.31 \\
FCN & 10.47 $\pm$ 3.77 & 1.76 $\pm$ 0.46 \\
Residual FCN & \textbf{7.64 $\pm$ 3.38} & \textbf{1.47 $\pm$ 0.43} \\
DCT & 12.10 $\pm$ 4.29 & 1.51 $\pm$ 0.64 \\
Modified ConvNext & 9.37 $\pm$ 4.40 & 1.52 $\pm$ 0.41 \\ \bottomrule
\end{tabular}
\end{table}

\begin{figure}[b]
    \centering
    \includegraphics[width=0.5\textwidth]{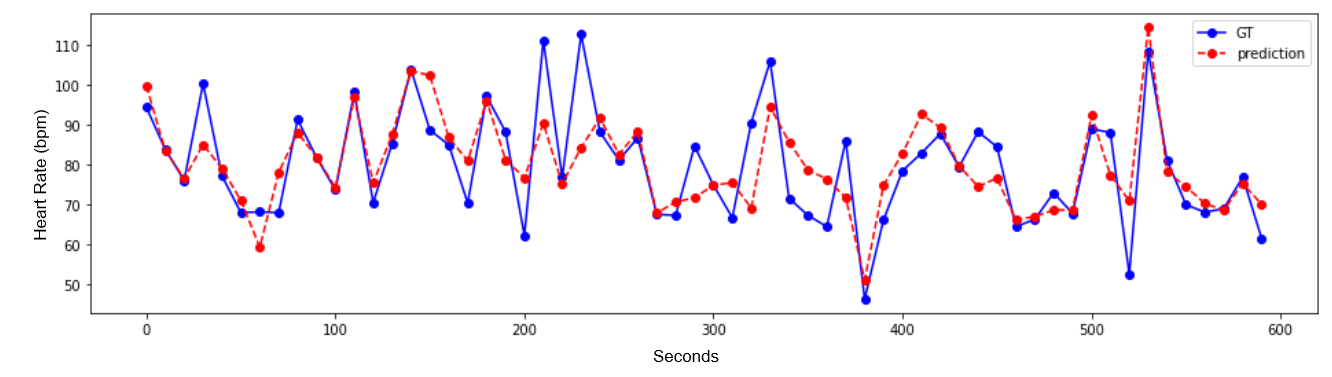}
    \caption{A plot of ground truth heart rate and its estimation by the best performing model (Residual FCN) on the MTHS dataset (MAE).}
    \label{fig:MTHS-pred-hr}
\end{figure}

\begin{figure}[ht]
    \centering
    \includegraphics[width=0.5\textwidth]{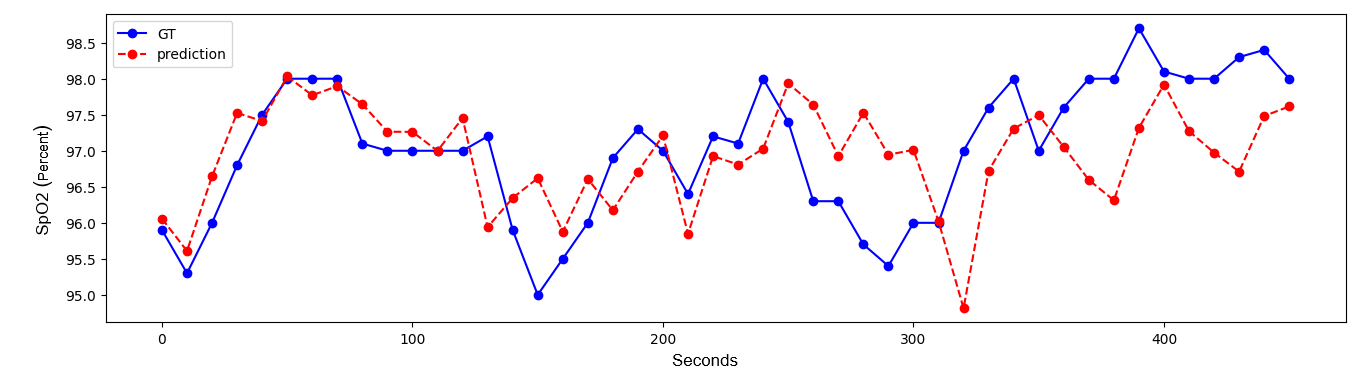}
    \caption{A plot of ground truth SpO2 and its estimation by the best performing model (Residual FCN) on the MTHS dataset (MAE).}
    \label{fig:MTHS-pred-spo2}
\end{figure}

In figures \ref{fig:MTHS-pred-hr} and \ref{fig:MTHS-pred-spo2}, sixty prediction samples are provided for both HR and SpO2 estimation along with their corresponding ground truth labels.

\subsection*{Comparison with other methods}
In this section, we undertake a comparative analysis, contrasting the performance of the proposed methods against alternative approaches across three distinct tasks: heart rate estimation on the PPG-DaLiA dataset, SpO2 and RR estimation on BIDMC.

\subsubsection*{\textbf{Heart Rate estimation on PPG-DaLiA}}

Table \ref{table:ppg-dalia}, demonstrates the performance of the proposed methods in this research and other competitive approaches. We use Leave One Subject Out (LOSO) cross-validation strategy to evaluate our proposed models on this dataset. To better compare and demonstrate the efficiency of our algorithms, we compute an efficiency score which is defined as:
\begin{center}
    $Efficiency = \dfrac{1}{ \#parameters \times MAE }$
\end{center}

In the evaluation of heart rate estimation on the PPG-DaLiA dataset, SpaMa and SpaMaPlus serve as baseline methods \cite{salehizadeh2015novel}, with SpaMaPlus consistently surpassing SpaMa across all sessions and the overall dataset, exhibiting MAE values between 6.31 and 24.06 bpm. Schäck et. al.  \cite{schack2017computationally}, while presenting higher MAE values (8.72 to 33.05 bpm), is outperformed by the CNN-based methods, CNN average and CNN ensemble, which demonstrate competitive performance with lower MAE values compared to SpaMa and Schaeck2017, with CNN ensemble generally outperforming CNN average.

The proposed methods (Base, FCN, Residual FCN, and DCT) consistently outshine the baseline methods and are comparable to or superior to the CNN-based methods. Notably, Residual FCN emerges as the most effective method for heart rate estimation, achieving the lowest MAE values ranging from 2.70 to 11.88 bpm across all sessions and the entire dataset. In-depth analysis reveals that the basic proposed method (Base) performs well, outperforming Schäck et. al, SpaMa, and SpaMaPlus. The Fully Convolutional Network (FCN) exhibits enhanced performance, and Residual FCN further refines the results, proving to be the most robust method. Additionally, the Discrete Cosine Transform (DCT) method, even without inverse transformation, competes favorably with MAE values ranging from 4.01 to 15.00 bpm, although it is surpassed by the Residual FCN in overall effectiveness. These findings underscore the efficacy of the proposed Residual FCN architecture in accurately estimating heart rates from physiological signals in the PPG-DaLiA dataset.

In our study, the Residual FCN stands out as our top-performing model, demonstrating strong performance. However, it is noteworthy that BeliefPPG surpasses Residual FCN in terms of both mean absolute error (MAE) and standard deviation (SD), despite the fact that the former has 2.6 times more parameters than our model. Compared to MH-Conv-LSTM our approach is a lot more efficient having 13x fewer parameters. This model has also utilized LSTM, which significantly increases inference time due to the sequential nature of its processing and the prolonged dependencies it introduces in the computational pipeline. NAS-PPG, while performing nearly the same in terms of MAE but the standard deviation is 4 times higher than ours. Moreover, it has 15x more parameters than our model.  This highlights the efficiency of our approach, showcasing that superior results can be achieved with a more streamlined and resource-efficient model. Our emphasis on efficiency positions our work as a compelling choice in the pursuit of effective and resource-conscious modeling. 

The proposed Residual FCN method demonstrates superior performance in heart rate estimation on the PPG-DaLiA dataset compared to baseline methods and other state-of-the-art approaches. The results suggest that the proposed Residual FCN architecture is effective in capturing complex relationships within the physiological signals, leading to more accurate heart rate predictions.

\begin{table*}[htbp]
\tiny
\centering
\caption{Session-wise evaluation results on the dataset PPG-DaLiA, achieved with proposed methods in this paper (Base, FCN, Residual FCN, and DCT). Results are given as MAE [bpm].}
\label{table:ppg-dalia}
\begin{tabularx}{\textwidth}{p{1.5cm}p{0.35cm}p{0.35cm}p{0.35cm}p{0.35cm}p{0.35cm}p{0.35cm}p{0.35cm}p{0.35cm}p{0.35cm}p{0.35cm}p{0.35cm}p{0.35cm}p{0.35cm}p{0.35cm}p{0.35cm}p{0.8cm}p{0.35cm}p{0.35cm}}
\toprule
 & S1  & S2  & S3 & S4 & S5 & S6& S7& S8& S9& S10 & S11 & S12 & S13 & S14& S15 & ALL& \#Param & Efficiency \\ \hline
SpaMa\cite{salehizadeh2015novel}  &11.86 & 14.75 & 9.53 & 17.2 & 39.28 & 16.78 & 15.88 & 15.2 & 17.19 & 9.08 & 21.63 & 12.63 & 9.5 & 10.73 & 12.23 & 15.56±7.5 & - & - \\
SpaMaPlus\cite{salehizadeh2015novel}   &8.86 & 9.67 & 6.4 & 14.1 & 24.06 & 11.34 & 6.31 & 11.25 & 16.04 & 6.17 & 15.15 & 12.03 & 8.5 & 7.76 & 8.29 & 11.06$\pm$4.8 & - & -\\
Schäck et. al.\cite{schack2017computationally} &33.05 & 27.81 & 18.49 & 28.82 & 12.64 & 8.72 & 20.65 & 21.75 & 22.25 & 12.6 & 21.05 & 22.74 & 27.71 & 12.05 & 16.4 & 20.45$\pm$7.1 & -& -\\
CNN average &8.45 & 7.92 & 5.96 & 7.86 & 18.97 & 13.55 & 5.16 & 11.49 & 10.65 & 6.07 & 9.87 & 9.95 & 5.25 & 5.85 & 5.25 & 8.82$\pm$3.8 & -& -\\
CNN ensemble &7.73 & 6.74 & 4.03 & \textbf{5.9} & 18.51 & 12.88 & 3.91 & 10.87 & \textbf{8.79} & \textbf{4.03} & 9.22 & 9.35 & 4.29 & 4.37 & \textbf{4.17} & 7.65$\pm$4.2 & 8.5M & 0.02\\ \hline
Base & 6.70  & 7.03  & 5.03 & 6.99 & 17.70 & 11.27& 4.04& 8.56& 11.18& 6.42 & 8.98 & 6.83 & 4.70 & 4.85& 6.59 & 7.79$\pm$3.36 & 29K& 4.43\\
FCN & 5.30  & 5.38  & 3.79 & 6.24 & 14.03 & 8.73& 3.15& 8.09& 10.64& 5.17 & 7.18 & 6.42 & 3.69 & 4.28& 5.24 & 6.49$\pm$2.82 &8K& 19.26\\
Residual FCN & \textbf{4.85}  & \textbf{4.69}  & \textbf{3.25} & 6.03 & \textbf{11.88} & \textbf{6.27}& \textbf{2.70}& \textbf{7.77}& 9.83& 4.12 & \textbf{6.15} & \textbf{6.26} & \textbf{2.94}  & \textbf{3.47}& 4.50 & \textbf{6.07$\pm$2.70} & 53K & 3.11\\
DCT(No Inverse) & 5.92  & 6.28  & 4.80 & 6.85 & 15.00 & 9.89& 4.01& 8.25& 11.24& 5.54 & 8.35 & 6.87 & 4.42 & 5.08& 6.22 & 7.25$\pm$2.84 & 3K & 45.98\\
BeliefPPG \cite{bieri2023beliefppg} & -  & -  & - & - & - & -& -& -& -&- & - & - & - & -& - & 3.57 $\pm$ 1.4 & 138K & 2.03 \\
BeliefPPG / viterbi \cite{bieri2023beliefppg} & -  & -  & - & - & - & -& -& -& -&- & - & - & - & -& - & 3.18 $\pm$ 1.3 & 138K & 2.28 \\
MH Conv-LSTM \cite{wilkosz2021multi}  & -  & -  & - & - & - & -& -& -& -&- & - & - & - & -& - & 6.28 $\pm$ 3.5  & 680K & 0.23 \\
NAS-PPG \cite{song2021ppg}  & -  & -  & - & - & - & -& -& -& -&- & - & - & - & -& - & 6.02$\pm$10.6  & 800K & 0.21\\

\bottomrule
\end{tabularx}
\end{table*}

\subsubsection*{\textbf{SpO2 estimation on BIDMC}}
In the following, the performance of the proposed models is compared against other methods. During the research process, no research was found reporting the HR results on BIDMC. Most existing research has focused on predicting the Respiratory Rate (RR), which will be provided in the next subsection. Table \ref{table:BIDMC_comparison_spo2} compares the performance of the proposed models on the task of SpO2 estimation with other approaches on the the BIDMC dataset.

\begin{table}[htbp]
\centering
\caption{Performance comparison of SpO2 estimation on BIDMC.}
\label{table:BIDMC_comparison_spo2}
\begin{tabular}{lllll}
\toprule
Model Name / Metric & MAE \\ \hline
Koteska et. al.\cite{koteska2022machine-b16}               & 1.45     \\
Shuzan et. al.\cite{shuzan2023machine}               & \textbf{0.98}     \\
\hline
Base                     & 3.56$\pm$0.72     \\
FCN                      & 3.4$\pm$0.91     \\
Residual FCN            & 3.67$\pm$0.81    \\
DCT (No inverse)        & 3.25$\pm$0.72    \\
Modified ConvNext        & 3.49$\pm$1.12    \\
\bottomrule
\end{tabular}
\end{table}

\subsubsection*{\textbf{RR Estimation on BIDMC}}
In this part, the performance of the proposed models is compared with similar methods for Respiratory Rate (RR) estimation. It should be noted that to evaluate our models, the KFold cross-validation technique was used with ten folds. The mean and standard deviation for all the metrics are reported across folds in Table \ref{table:MTHS-MAE-rr}. According to the table, the Residual FCN model proposed in this research outperforms all other proposed methods, as well as other state-of-the-art methods except for \cite{shuzan2023machine}.

\begin{table}[htbp]
\centering
\caption{Performance comparison of RR estimation on BIDMC.}
\label{table:MTHS-MAE-rr}
\begin{tabular}{lll}
\toprule
Model Name  & MAE  & RMSE  \\ \hline
Base                    & 2.65$\pm$\textbf{0.74} & 13.83$\pm$7.11     \\
FCN                      & 1.94$\pm$0.79 & 8.62$\pm$7.16   \\
Residual FCN             & \textbf{1.49}$\pm$0.91 & \textbf{5.10}$\pm$\textbf{4.93}    \\
DCT                      & 2.29$\pm$\textbf{0.74} & 11.55$\pm$9.97   \\
Modified-Convnext                & 1.51$\pm$0.88& 5.71$\pm$5.57  \\
\hline
Shuzan\cite{shuzan2023machine} & \textbf{0.89} &\textbf{1.41} \\
Jarchi\cite{jarchi2018accelerometry} & 2.56 & - \\
Bian \cite{bian2020respiratory} DL &3.8±0.5 &-  \\
Bian \cite{bian2020respiratory} SQF & 2.6±0.4 & - \\
Aqajari\cite{aqajari2021end}& 	1.9±\textbf{0.3} & - \\
\bottomrule
\end{tabular}
\end{table}

\section{Ablation Study}
In this section, a series of experiments were conducted to assess the effect of using different training loss functions on the final error and also the effect of using different color channels for estimating the heart rate. 

\subsection*{Loss Functions}
Four different loss functions were selected for training the proposed architectures. These loss functions include Mean Absolute Error (MAE), Mean Squared Error (MSE), huber, and log-cosh. To make the training process faster, random splitting of the train, test and validation sets are taken instead of folded cross validation. The BIDMC dataset is split into three sets of train, validation, and test set with slice size of 0.8, 0.04, and 0.16, respectively, seeded with the value of 1400. Tables \ref{table:BIDMC-MAE-hr-ablation} and Table \ref{table:BIDMC-MAE-spo2-ablation} show the heart rate and SpO2 estimation results on BIDMC dataset with different loss functions. Our experiments generally show that the MAE and log-cosh losses are good choices. However, it is also observed that a single loss function that works best for a specific architecture does not necessarily work best on other architectures. 

\begin{table}[h]
\centering
\caption{Heart rate estimation results on the BIDMC test set (MAE). The results are reported for all proposed architectures with different loss functions.}
\label{table:BIDMC-MAE-hr-ablation}
\begin{tabular}{lllll}
\toprule
Model Name / Loss Function & MSE  & MAE  & Huber & Log-cosh \\ \hline
Base                     & 2.33 & 2.77  & 2.20  & 2.85     \\
FCN                      & 1.91 & 1.86  & 1.94  & 2.01     \\
Residual FCN            & 1.38 & 1.44 & 1.43 & \textbf{1.36}     \\
DCT (w/o inversion)        & 3.08 & 2.02 & 2.24  & 2.21    \\ \hline
Mean        & 2.18 & 2.02 & \textbf{1.95}  & 2.1    \\
\bottomrule
\end{tabular}
\end{table}

\begin{table}[h]
\centering
\caption{Spo2 estimation results on BIDMC test set (MAE). The results are reported for all proposed architectures with different loss functions.}
\label{table:BIDMC-MAE-spo2-ablation}
\begin{tabular}{lllll}
\toprule
Model Name / Loss Function & MSE  & MAE  & Huber & Log-cosh \\ \hline
Base                     & 2.25 & 1.94  & 2.04  & 1.97     \\
FCN                      & 2.08 & 1.85  & 1.93  & 1.98     \\
Residual FCN            & 1.20 & 1.02 & 1.07 & \textbf{1.00}     \\
DCT (No inverse)        & 2.25 & 2.12 & 2.16  & 2.08    \\ \hline
Mean        & 1.95 & \textbf{1.73} & 1.8  & \textbf{1.76} \\
\bottomrule
\end{tabular}
\end{table}

MTHS is split into three sets of train, validation, and test set with slice size of 0.68, 0.12, and 0.2, respectively, seeded with the value of 1400. Tables \ref{table:MTHS-MAE-hr-ablation} and \ref{table:MTHS-MAE-spo2-ablation} show the heart rate and SpO2 estimation results on BIDMC dataset with different loss functions.

\begin{table}[htbp]
\centering
\caption{Heart rate estimation results on MTHS test set (MAE). The results are reported for all proposed architectures with different loss functions.}
\label{table:MTHS-MAE-hr-ablation}
\begin{tabular}{lllll}
\toprule
Model Name / Loss Function & MSE  & MAE  & Huber & Log-cosh \\ \hline
Base                     & 10.92 & 10.75  & 9.95  & 10.62     \\
FCN                      & 8.98 & 8.42  & 8.58  & 8.35     \\
Residual FCN             & 8.33 & \textbf{6.59} & 6.92  & 6.97     \\
DCT                     & 10.09 & 9.02 & 9.80  & 8.67    \\\hline
Mean        & 9.58 & 8.70 & 8.81  & \textbf{8.65} \\
\bottomrule
\end{tabular}
\end{table}

\begin{table}[htbp]
\centering
\caption{SpO2 estimation results on MTHS test set (MAE). The results are reported for all proposed architectures with different loss functions.}
\label{table:MTHS-MAE-spo2-ablation}
\begin{tabular}{lllll}
\toprule
Model Name / Loss Function & MSE  & MAE  & Huber & Log-cosh \\ \hline
Base                    & 1.63 & 1.43  & 1.50  & 1.54     \\
FCN                      & 1.72 & 1.64  & 1.40  & 1.46     \\
Residual FCN             & 1.36 & \textbf{1.24} & 1.34  & 1.32     \\
DCT                      & 1.47 & 1.36 & 1.41  & 1.41   \\\hline
Mean        & 1.55 & \textbf{1.42} & \textbf{1.41}  & \textbf{1.43} \\
\bottomrule
\end{tabular}
\end{table}

\subsection*{Color Channel for HR Estimation}
The effect of using different color channels as the input PPG signal is examined in this experiment on the MTHS dataset. The best-performing model is trained with log-cosh loss, each time with a different color channel as input PPG signal. 10-fold cross-validation is applied on the MTHS dataset. The mean and standard deviation of MAE are reported. Table \ref{table:color-channel-mths-ablation} demonstrates these results. According to the experimental results, using the green channel yielded the best results.

\begin{table}[htbp]
\centering
\caption{The results of HR estimation for different color channels on the MTHS dataset. The results are reported with 10 fold cross validation setup.}
\label{table:color-channel-mths-ablation}
\begin{tabular}{lllll}
\toprule
Model/ Channel  & Red  & Green  & Blue & All \\ \hline
Residual FCN             & 7.97$\pm$4.18 & \textbf{7.54$\pm$3.39} & 9.93$\pm$4.14  & 8.33$\pm$4.45     \\
\bottomrule
\end{tabular}
\end{table}

\section{Deployment on Smart Phone}

\begin{figure}[h]
    \centering
    \includegraphics[width=\textwidth/2]{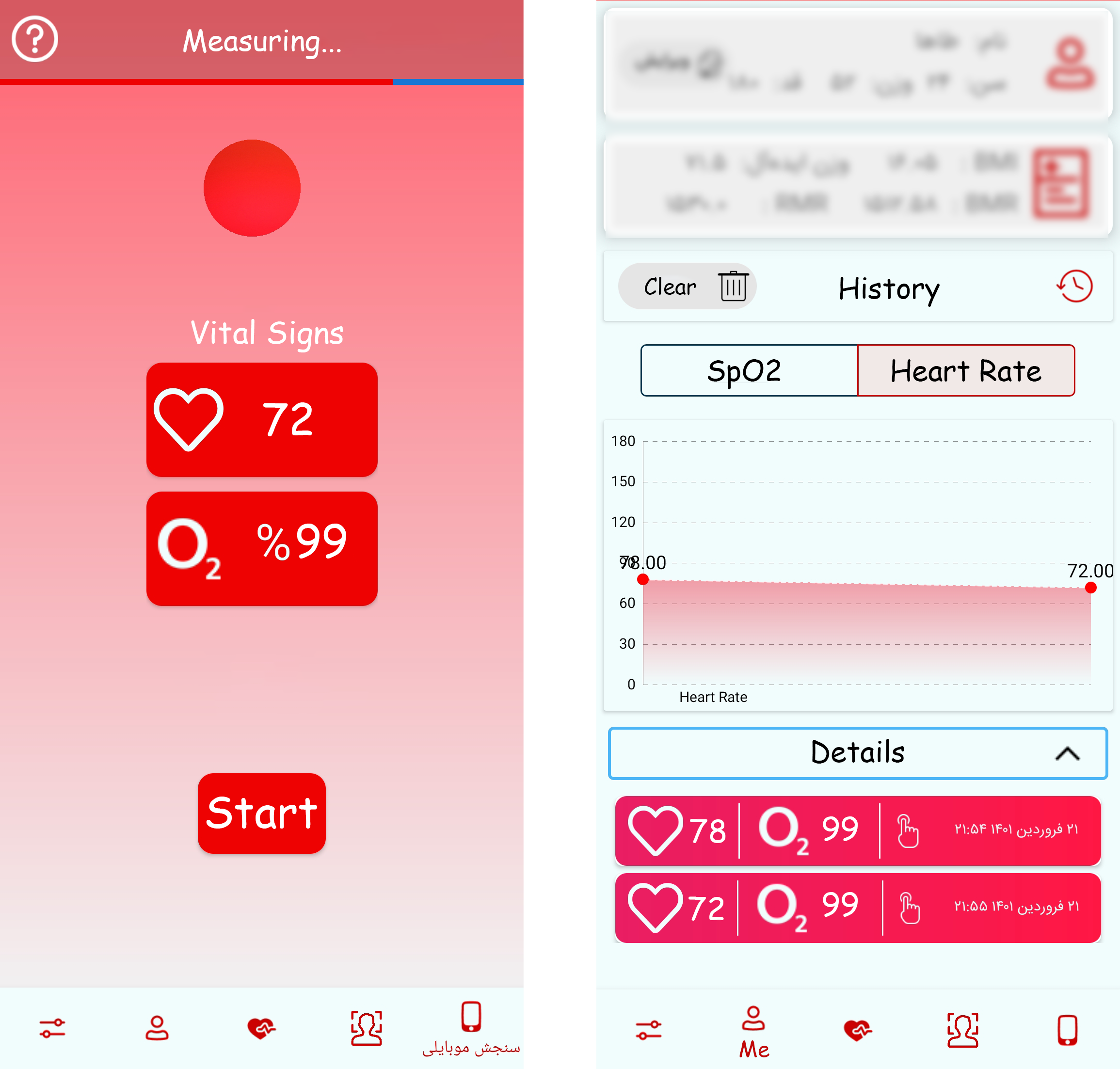}
    \caption{Screenshots of the developed application showing the measurement and profile sections. MEDVSE is deployed on the android application for vital sign estimation. The predicted vital signs are displayed to the user and updated every 10 seconds.}
    \label{fig:application-screenshot}
\end{figure}

Figure \ref{fig:application-screenshot} shows some screenshots of the developed application. The application is developed for the android platform and is compatible with version 5 (Lollipop) and above. The deep learning models for heart rate and SpO2 estimation have been converted to TFLite \cite{tflite-b17} and integrated into the application. The user is requested to cover the camera while the flashlight is on at the running time. After the finger is detected, the estimation process starts. The application captures video frames at 30 fps. Every 10 seconds, a mean signal of red, green, and blue channels is computed and given to the deep learning model to infer the vital signs.

\section{Discussion}
This research explored efficient deep-learning architectures for vital sign estimation using smartphone cameras. The results demonstrate the promise of fully convolutional networks for this application. The proposed Residual FCN architecture has showcased state-of-the-art performance in heart rate, SpO2, and RR estimation on benchmark datasets, surpassing many previous CNN-based methods. Notably, it exhibits relatively high parameter efficiency and reduces computational overhead by omitting batch normalization layers. With these advantages, including a model size lower than 1 MB, the architecture contributes to the feasible real-time estimation of multiple vital signs, even on mid-range smartphones.

Additionally, the DCT model shows that working in the frequency domain can also enable compact yet accurate models. By filtering the input signal to the relevant frequency bands (this is integrated into the model), the DCT model achieves competitive performance to standard CNNs with much fewer parameters and under 1 million FLOPs.

Our introduction of the public MTHS dataset captured using smartphone cameras helps address the shortage of PPG data for mobile health applications of deep learning. With further augmentation, this dataset could facilitate advances in personalized models that account for differences across age, gender, skin type, etc. Unsupervised domain adaptation techniques may also leverage the paired phone-oximeter data we provide. A possible future work here is to collect SpO2 and HR data from people with respiratory-related diseases; Since our dataset only contains the data of healthy subjects.

Future work includes further model optimization and compression to maximize efficiency without sacrificing accuracy. While focused on PPG signals, exploring multi-modal sensor inputs like accelerometers could make the models more robust to motion artifacts. Though outside our current scope, jointly predicting related variables like heart rate variability could provide richer clinical insights. Testing performance across devices and demographics may reveal generalizability limitations to guide additional data collection for fine-tuning. 

Overall, through systematic experiments and comparisons, our work demonstrates the potential of deploying highly efficient deep neural networks on ubiquitous smartphones to enable continuous physiological monitoring without dedicated wearable devices. The clinical acceptability of such mobile vital sign estimators could be established through more rigorous validation. This could open up avenues for scalable remote patient care as well as personalized diagnostics and interventions.

\section{Conclusion and Future Work}
Due to the necessity of regular monitoring of vital signs, especially in the elderly, one can take advantage of smartphones for estimating and monitoring vital signs with acceptable accuracy. This research proposed an end-to-end, real-time, deep learning model for smartphone-based vital sign estimation. Our method has fewer parameters than previous methods and only needs a normalization step to be applied to the signal which is integrated into the model. The proposed model can be deployed even on low-end devices due to the low computational complexity. There is also a public dataset of smartphone videos named MTHS provided which contains extracted PPG signals from 62 distinct patients with their corresponding ground truth HRs and SpO2s. A series of experiments were performed on semi-supervised learning methods during this research. The employed approach for training the model was semi-supervised, wherein synthesised PPG signals and their corresponding heart rate labels were generated automatically. However, the model's performance did not meet the expected standards to report the results. Even though fully convolutional structures can have lower parameters and better performance than the ones with fully connected prediction heads, the FLOPs required by FCN models are higher. Given the lack of good quality and quantity data for smartphone-based vital sign estimation, the authors believe there is a strong potential for unsupervised learning methods that can be used in this field.

\section*{Acknowledgment}
Part of this research was supported by Iran's National Elites Foundation. Special Thanks to Vahid Bastani, Mahdis Habibpour, Ali Farvardin, and Changiz Javadian for their kind support and help during this project.

\bibliographystyle{ieeetr}
\bibliography{bibliography}

\end{document}